\documentclass[prc,twocolumn,showpacs,amsmath,amssymb]{revtex4}
\usepackage{graphicx}
\usepackage{dcolumn}
\def\thalf{{\textstyle{\frac{1}{2}}}}

\begin{document}
\preprint{NUC-MINN-04/1-T}
\title{Properties of the $\omega$ Meson at Finite Temperature and 
Density}
\author{A. T. Martell and P. J. Ellis}
\affiliation{School of Physics and Astronomy,
University of Minnesota, Minneapolis, MN 55455, USA} 
\date{\today}

\begin{abstract}
The mass shift, width broadening, and spectral density for the $\omega$ 
meson in a heat bath of nucleons and pions is calculated using a general 
formula which relates the self-energy to the forward scattering amplitude.
We use experimental data to saturate the scattering amplitude at low 
energies with resonances and include a background Pomeron term, while at 
high energies a Regge parameterization is used. The peak of the spectral 
density is little shifted from its vacuum position, but the width is 
considerably increased due to collisional broadening. At normal nuclear 
matter density and a temperature of 150 MeV the spectral density of the 
$\omega$ meson has a width of 140 MeV. Zero temperature nuclear matter is 
also discussed.
\end{abstract}
\maketitle

The modification of the free space properties of a vector meson
in hadronic or nuclear matter is an important problem which has
attracted much attention, see the reviews of Ref. \cite{rev}.
Many works have relied on effective Lagrangians, however
in Ref. \cite{us} (hereafter I) we adopted as model-independent an 
approach as possible by using experimental data to construct the amplitude 
for vector mesons scattering from pions and nucleons.
The low energy region was described in terms of resonances plus background,
while at high energies a Regge model was employed. Using this amplitude 
the in-medium self-energy of the 
$\rho$ and $\omega$ mesons was calculated at finite temperature and 
density using the leading term of the exact self-energy expansion 
\cite{je}. This requires that only single scatterings be important
which was found to be justified in I by comparison with results from 
ultrarelativistic molecular dynamics (UrQMD) calculations. This procedure was 
satisfactory for the $\rho$ meson and gave results consistent with those 
obtained by Rapp \cite{rap} by considering medium modifications of the 
pions comprising the meson. Less satisfactory was the $\omega$ meson case 
because little data was available for the decay of nucleon resonances in the
$\omega N$ channel. Therefore in I two extreme models were adopted: a
two-resonance model using the data of Manley and Saleski \cite{man} and an 
$\omega\rightarrow\rho$ model which assumed that the resonance decays in 
the $\omega N$ and $\rho N$ channels were essentially the same. In the 
meantime much better resonance 
data in the $\omega N$ channel has become available with the analysis of 
Shklyar, Penner and Mosel \cite{pm}. The purpose of the present paper is 
to use this new data to provide a more reliable $\omega$ self-energy than 
the two extreme models adopted in I. 

First we consider the low energy regime.
We assume that the $\omega$ self-energy is dominated by scattering from 
the pions and nucleons present in the heat bath, as was justified in I
by comparison with UrQMD results. Focussing on the latter
we briefly outline the formalism for constructing the  $\omega N$ amplitude,
more details are to be found in I. We adopt the two-component duality 
approach due to Harari \cite{har} (see also Collins \cite{col}) which states 
that while ordinary Reggeons are dual to $s$-channel
resonances, the Pomeron is dual to the background upon which the resonances 
are superimposed. We write the forward scattering amplitude in the rest
frame of the heat bath 
\begin{eqnarray}
f_{\omega N}(E_\omega)&=&\frac{\sqrt{s}}{2q_{\rm cm}m_N}\sum_R 
W^R_{\omega N}\frac{\Gamma_{R\rightarrow\omega N}}
{M_R-\sqrt{s}-\thalf i\Gamma_R}\nonumber\\
&&-\frac{q_{\rm cm}r^{\omega N}_P}{4\pi m_N \sqrt{s}}
\frac{(1+\exp^{-i\pi\alpha_P})}{\sin\pi\alpha_P}s^{\alpha_P}\;. \label{lef}
\end{eqnarray}
Here the first term involves a sum over a series of Breit-Wigner 
resonances of mass $M_R$ and total width $\Gamma_R$ (replacement of this 
non-relativistic form by the relativistic expression has a negligible 
effect on the results). The second term is the Pomeron background 
contribution discussed below. In the usual notation $\sqrt{s}$ is the total 
energy which is related to the energy of the $\omega$ meson by
$E_\omega-m_\omega=[s-(m_\omega+m_N)^2]/(2m_N)$,
where $m_\omega$ and $m_N$ are the omega and nucleon masses.
In Eq. (\ref{lef}) $q_{\rm cm}$ denotes the magnitude of the c.m. momentum 
and the statistical averaging factor  
$W^R_{\omega N}=(2s_R+1)/6$, where $s_R$ is the spin of the resonance.
Since we are averaging over all spin directions we 
shall not distinguish longitudinal and transverse polarizations. 
Also $\Gamma_{R\rightarrow\omega N}$ in Eq. (\ref{lef}) represents 
the partial width for the resonance decay into the $\omega N$ channel.
If we denote the c.m. momentum at resonance by $q_{\rm cm}^R$, then
for $q_{\rm cm}\geq q_{\rm cm}^R$ we use the value obtained from the total 
width and the branching ratio on resonance. However the threshold behavior 
of the partial width is known and we incorporate this for
$q_{\rm cm}\leq q_{\rm cm}^R$ by replacing
$\Gamma_{R\rightarrow\omega N}$ by $\Gamma_{R\rightarrow\omega N}
(q_{\rm cm}/q_{\rm cm}^R)^{2l+1}$, where $l$ is the relative angular 
momentum between the $\omega$ and the nucleon. Since the total width is 
the sum of the partial widths in principle this dependence should be 
incorporated in $\Gamma_R$, but this is impractical as there are many 
decay channels open so we simply take $\Gamma_R$ to be a constant.

The above-threshold resonances included in our calculation are listed in 
Table \ref{tab1}. The first five entries are taken from the fit labelled 
C-p-$\pi$+(5/2) by Shklyar, Penner and Mosel \cite{pm} which extends to 
2 GeV. We also include the $N(2190)$ strength taken from  Manley 
and Saleski \cite{man};  Vrana, Dytman and Lee \cite{lee} report
a roughly similar width and branching ratio for this resonance.
It is also necessary to include subthreshold resonances since they make a 
significant contribution. We include the $N(1520)\:D_{13}$ and the  
$N(1535)\:S_{11}$, as well as the smaller contributions from the
$N(1440)\:P_{11}$, the $N(1650)\:S_{11}$ and the $N(1680)\:F_{15}$ resonances.
In order to estimate the widths we assume that the vector dominance model 
is valid, even though it is better suited to high energies. This allows us 
to relate the photon and $\omega$ widths. Specifically, since these resonances 
are close to the $\omega N$ threshold, we can write for each of them 
$\Gamma_{\omega N}=q_{{\rm cm}}\gamma_{\omega N}$
and $\Gamma_{\gamma N}=k_{{\rm cm}}\gamma_{\gamma N}$,
where $k_{{\rm cm}}$ is the $\gamma N$ c.m. momentum.
Then vector dominance gives 
\begin{equation}
\gamma_{\gamma N} = 4\pi \alpha \frac{1}{g^2_{\rho}}\Biggl ( 1 +
\frac{g^2_{\rho}}{g^2_{\omega}} \Biggr ) \gamma_{\omega N} \;,
\end{equation}
where $\alpha$ is the fine structure constant. For the coupling to the 
photon we take $g_\rho^2/4\pi=2.54$ and $g_\rho^2/g_\omega^2=1/8$.
The value of $\gamma_{\gamma N}$ can be deduced from the decay width and 
the photon branching ratio of the resonances \cite{pdg}. 

\begin{table}[t]
\begin{ruledtabular}
\caption{Baryon resonances included in the $\omega N$ amplitude}
\begin{tabular}{|l|ccr|}
&Mass&Width &
$\omega N$ Branching Ratio\\Resonance&(GeV)&(GeV)&(\%)\\ \hline
$N(1710)\:P_{11}$&1.753&0.534&19.9\\
$N(1720)\:P_{13}$&1.725&0.267&0.8\\
$N(1900)\:P_{13}$&1.962&0.700&9.6\\
$N(1950)\:D_{13}$&1.927&0.855&47.0\\
$N(2000)\:F_{15}$&1.981&0.361&2.2\\
$N(2190)\:G_{17}$&2.127&0.547&49.0\\ 
\end{tabular}
\label{tab1}
\end{ruledtabular}
\end{table}

The high energy forward scattering amplitude is known \cite{don} 
to be well approximated by the Regge form
\begin{equation}
f_{\omega N}(E_\omega)=-\frac{q_{\rm cm}}{4\pi m_N\sqrt{s}}\sum_i
\frac{1+\exp^{-i\pi\alpha_i}}{\sin\pi\alpha_i}r^{\omega N}_is^{\alpha_i}\;. 
\label{hef}
\end{equation}
We shall consider a Pomeron term $P$ and a Regge term $P'$. Since the 
different isospin structure of the $\omega$ and the $\rho$
is expected to be insignificant at high energy, we adopt the same
parameterization for the $\omega N$ and $\rho N$ scattering amplitudes 
as in I. Specifically the intercepts are $\alpha_P=1.093$ and 
$\alpha_{P'}=0.642$ with residues $r_P^{\omega N}=11.88$ and 
$r_{P'}^{\omega N}=28.59$. The units are such
that with energies in GeV the total cross section is given in mb; 
specifically the optical theorem gives 
$\sigma=4\pi{\rm Im}f_{\omega N}/p$ where the momentum in the rest frame 
of the heat bath $p=q_{\rm cm}\sqrt{s}/m_N$. The parameters for the 
Pomeron given here are also used for the background term in Eq. (\ref{lef}). 
Note that if the Pomeron intercept $\alpha_P$ were exactly 1, the Pomeron 
amplitude would be pure imaginary.

\begin{figure}[t]
\includegraphics[width=8truecm]{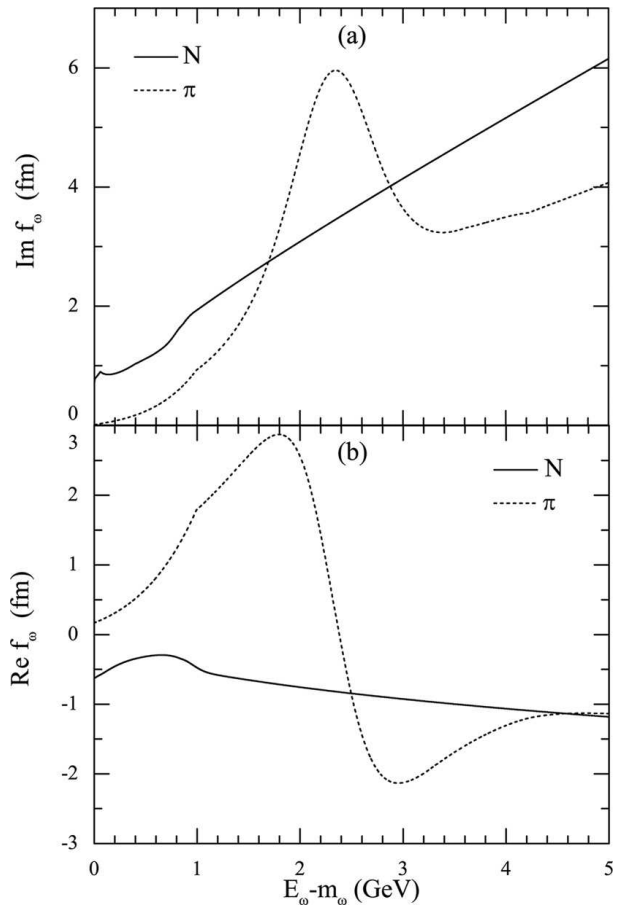}
\caption{(a) the imaginary and (b) the real part of the amplitude 
for $\omega N$ 
scattering (solid line) and $\omega\pi$ scattering (dashed line).} 
\label{fig:one}
\end{figure}

Because of the kinematics the resonance region 
ends at $E_\omega-m_\omega\sim1$ GeV and the amplitude is smoothly 
matched onto the Regge part at approximately this point. 
The real and imaginary parts of $f_{\omega N}$ constructed in this manner 
are indicated by the solid curves in  Fig. \ref{fig:one}. Since the
low energy part contains a number of overlapping resonances the structure 
is washed out. We also indicate by dashed curves in  Fig. \ref{fig:one} 
the corresponding results 
for $f_{\omega\pi}$ taken from I for which the single $b_1(1235)$ resonance
employed is clearly visible (note that due to kinematics the resonance 
region ends at $E_\omega-m_\omega\sim4$ GeV for the $\omega\pi$ system).
Our result for the imaginary part of $f_{\omega N}$ in Fig. \ref{fig:one}(a) 
can be compared 
with that obtained by Sibirtsev, Elster and Speth \cite{sib} using 
data for $\omega$ photoproduction from nuclei and employing the eikonal 
approximation and vector meson dominance. There is good qualitative 
agreement, while quantitatively our values are 20--30\% lower than theirs.
As these authors point out the real part at threshold is quite uncertain,
even as regards sign. At the highest energy we consider the ratio of the real 
to the imaginary part agrees quite nicely with the $\pi N$ 
data \cite{sib,fol}. However this is to be expected since in I the high 
energy Regge behavior was fixed by using the charge-averaged $\pi N$ data.

For an $\omega$ meson scattering from a hadron $a$ in the medium 
the retarded self-energy on shell can be written \cite{us,je,ele} as a single 
integral. For the case that $a$ is a boson the result is
\begin{eqnarray}
\Pi_{\omega a}(p) &=& - \frac{m_\omega m_a T}{\pi p} \int\limits_{m_a}^{\infty}
d\omega f_{\omega a}\left(\frac{m_\omega \omega}{m_a}\right)\nonumber\\
&&\qquad\qquad\times\ln\left[\frac{1-\exp(-\omega_+/T)}{1-\exp(-\omega_-/T)}
\right] \;.
\end{eqnarray}
Here $\omega^2=m_a^2+k^2$ and $\omega_{\pm} = (E \omega \pm pk)/m_\omega$
with $E^2=m_\omega^2+p^2$.
If $a$ is a fermion $\omega_{\pm}$ has an additional chemical potential 
contribution $-\mu$ and the argument of the logarithm becomes
$[1+\exp(-\omega_-/T)]/[1+\exp(-\omega_+/T)]$.
The total self-energy is given by summing over all target species and 
including the vacuum contribution 
\begin{equation}
\Pi_{\omega}^{\rm tot}(E,p) =\Pi_{\omega}^{\rm vac}(M) + 
\Pi_{\omega \pi}(p) + \Pi_{\omega N}(p)\;.
\end{equation}
Here the vacuum part of $\Pi$ can only depend on the invariant mass,
$M = \sqrt{E^2 - p^2}$, whereas the matter parts can in principle depend on
$E$ and $p$ separately.  However, in the approximation we are
using the scattering amplitudes are of necessity evaluated on
the mass shell of the $\omega$ meson.  This means that the matter parts
only depend on $p$ because $M$ is fixed at $m_{\omega}$.  
The dispersion relation is determined from the poles of the propagator
with the self-energy evaluated on shell, {\it i.e.} $M=m_{\omega}$, giving
\begin{equation}
E^2 = m_{\omega}^2 + p^2 +\Pi_{\omega}^{\rm tot}(p)\;.
\end{equation}
Since the self-energy has real and imaginary parts so does
$E(p) = E_R(p) -i \Gamma(p)/2$.  The width is given by 
\begin{eqnarray}
&&\Gamma(p) = - {\rm Im}\Pi_{\omega}^{\rm tot}(p)/E_R(p) \; ,\quad{\rm with}\\
&&2E_R^2(p) = p^2+m_{\omega}^2+{\rm Re}\Pi_{\omega}^{\rm tot}(p)\nonumber\\
&&\qquad+\sqrt{[p^2+m_{\omega}^2+{\rm Re}\Pi_{\omega}^{\rm tot}(p)]^2  
+[{\rm Im}\Pi_{\omega}^{\rm tot}(p)]^2}   \, . 
\end{eqnarray}
In vacuum the width
$\Gamma_{\omega}^{\rm vac}=-{\rm Im}\Pi_{\omega}^{\rm vac}/m_{\omega}$ 
is 8.4 MeV. We define the mass shift to be
\begin{equation}
\Delta m_{\omega}(p) = \sqrt{m_{\omega}^2+{\rm Re}\Pi_{\omega}^{\rm tot}(p)} 
- m_{\omega} \;. 
\end{equation}

\begin{figure}[t]
\includegraphics[width=8truecm]{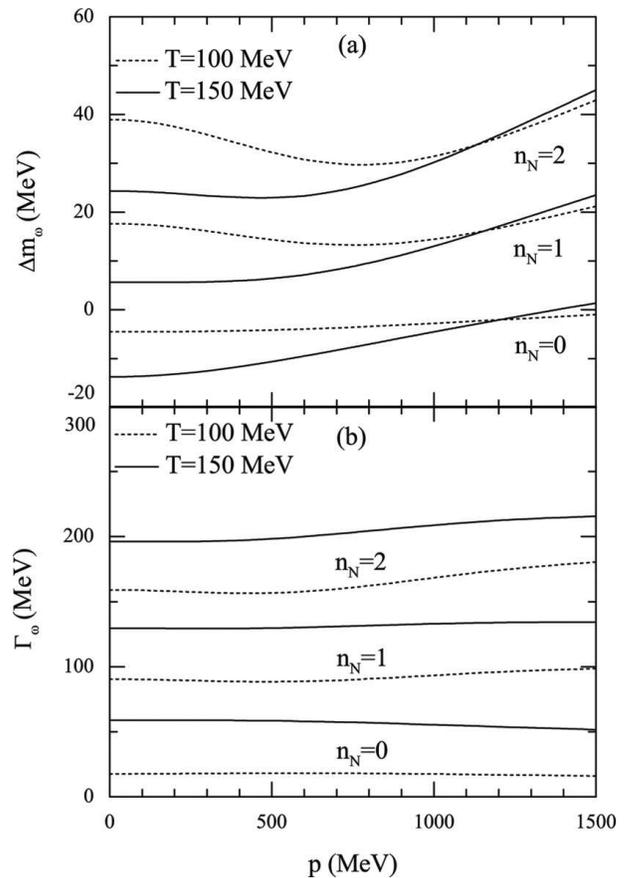}
\caption{(a) the  mass shift and (b) the width of the $\omega$ meson  
as a function of 
momentum $p$. Results are shown for nucleon densities of $0,\ 1$ 
and $2$ in units of equilibrium nuclear matter density and temperatures of 
100 and 150 MeV.} \label{fig:two}
\end{figure}

The $\omega$ meson mass shifts and widths are shown as a function of 
momentum in Fig. \ref{fig:two} for two temperatures and nucleon densities
$n_N=$0, 1 and 2 in units of equilibrium nuclear matter 
density $n_0$ (0.16 nucleons/fm$^3$). At zero nucleon density only
$f_{\omega\pi}$ is required so the results are the same as in I. In 
particular the mass shift, $\Delta m_\omega$, in Fig. \ref{fig:two}(a) 
is small and negative. It becomes positive when nucleons are introduced;
at large momentum $p$ the mass shifts are quite similar to the 
$\omega\rightarrow\rho$ model of I, however for $p=0$ the shifts are 
larger than given by either model of I. Notice that higher temperatures
lead to a smaller mass shift. However the values of $\Delta m_\omega$ are
at most a few tens of MeV, which is small in comparison to the vacuum
mass. Rapp \cite{rap}, who considers medium modifications of the pions 
comprising the omega, reports a negligible mass shift. For reference in 
zero-temperature nuclear matter, where pions do not contribute, the 
mass shift is approximately +30 MeV ($p=0$). There are a wide range of 
estimates in the literature. Post and Mosel \cite{post} obtain 20 MeV with 
a resonance model, Lutz, Wolf and Friman \cite{fri} find 70 MeV in a 
coupled channel approach and M\"uhlich, Falter and Mosel \cite{muh} 
with their adopted ${\rm Re}f_{\omega N}$ quote $-35$ MeV. A mass shift 
of much larger magnitude has been found in the chiral approach of
Klingl, Waas and Weise \cite{kling}. Large mass shifts have also been 
found using QCD sum rules \cite{rev,zsch}, but these are tailored to the 
small distance behavior whereas, as Eletsky and Ioffe \cite{ei} have 
pointed out, the self-energy is determined by meson-nucleon scattering 
at relatively large distances of order 1 fm; see also Ref. \cite{mal}.

Our results for the width $\Gamma_\omega$ are given in 
Fig. \ref{fig:two}(b). Note that the widths given here are defined to be
in the rest frame of the thermal system. As we have remarked, at $n_N=0$
the results are the same as in I. It was pointed out there that the 
calculated width of about 50 MeV at $T=150$ MeV is in agreement 
with the value of Schneider and Weise \cite{sch} and  a little larger 
than given by Haglin \cite{hag}. Alam et al. \cite{alam} using the Walecka 
model achieve only half of our value, but they find that the magnitude 
increases very steeply with temperature. Turning to finite values of $n_N$, 
our results are rather insensitive to the momentum $p$ and are intermediate 
between the two-resonance and $\omega\rightarrow\rho$ models discussed in I.
For a temperature of 150 MeV and $n_N=1$, $\Gamma_\omega=130$ MeV, an 
enhancement of the vacuum width by a factor of 15. This is in line with 
Rapp's estimate \cite {rap} of a factor of 20 at a slightly higher 
temperature of 180 MeV. These values are somewhat smaller than the 
results Riek and Knoll \cite{riek} obtained 
with self-consistent coupled Dyson equations. For zero-temperature nuclear 
matter we find the width due to collisional broadening to be 75 MeV. 
A similar result was obtained by Riek and Knoll \cite{riek}. On the other 
hand smaller values $\sim40$ MeV were found by Post and Mosel \cite {post},
by Lutz, Wolf and Friman \cite{fri} from a relativistic coupled 
channels approach and
by M\"uhlich, Falter and Mosel \cite{muh} in a transport model.

The rate of dilepton production is directly proportional to the
imaginary part of the photon self-energy \cite{mt,w} which is
itself proportional to the imaginary part of the $\omega$ meson
propagator because of vector meson dominance \cite{gs,gk}.
\begin{equation}
E_+ E_- \frac{dR}{d^3p_+ d^3p_-} \propto
\frac{-{\rm Im} \Pi_{\omega}^{\rm tot}}{[M^2 - m_{\omega}^2 
-{\rm Re} \Pi_{\omega}^{\rm tot}]^2+[{\rm Im}\Pi_{\omega}^{\rm tot}]^2}\;,
\end{equation}
where, as before, $M$ is the invariant mass.
Since the vacuum decay of the $\omega$ into three pions is  
complicated, while the width is tiny, we simply treat 
Im$\Pi_\omega^{\rm vac}$ as a constant
except for the application of a non-relativistic phase space factor
$[(M^2-9m_\pi^2)/(m_\omega^2-9m_\pi^2)]^2$ from threshold to $M=m_\omega$.
A possible real vacuum contribution is ignored.

\begin{figure}[t]
\includegraphics[width=8truecm]{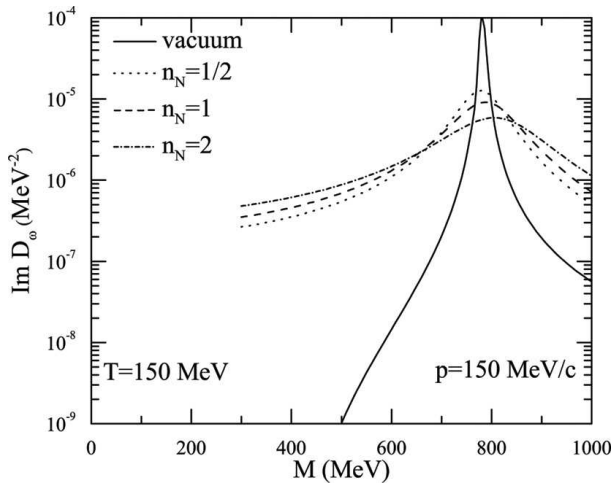}
\caption{The imaginary part of the $\omega$ meson propagator as a
function of invariant mass for a momentum of 300 MeV/c and a temperature of 
150 MeV. Results are shown for the vacuum and for nucleon densities of 
$\thalf ,\ 1$ and $2$ in units of equilibrium nuclear matter density.}
 \label{fig:three}
\end{figure}

The imaginary part of the propagator, proportional to the spectral density, 
is plotted as a function of $M$ in Fig. \ref{fig:three} for
a temperature of 150 MeV. Pions alone have a small effect on the spectral 
density so we display results at $n_N=\thalf,\ 1$ and $2n_0$.
These parameters are characteristic of the final stages
of a high energy heavy ion collision.  As seen from Fig. \ref{fig:three} there 
is little change in the position of the peak, but the spectral density 
is greatly broadened. For $n_N=1$ the width of the $\omega$ peak (full width, 
half maximum) is 140 MeV. This is quite similar to the $\omega\rightarrow\rho$ 
model in I and to the results of Rapp \cite{rap}.

In summary, the in-medium properties of the $\omega$ meson found in I 
have been updated by employing the $\omega N$ resonance analysis of
Ref. \cite{pm}. Taking as a reference point a temperature of 150 MeV,
equilibrium nuclear matter density and zero momentum, the $\omega$ 
mass shift of 6 MeV was negligible. However, the width of 130 MeV 
represented a considerable increase from the vacuum value (8.4 MeV). 
Thus the spectral density, which determines dilepton production
from this channel, was greatly broadened. These values are reasonably 
consistent with the bulk of the estimates in the literature. In particular 
the agreement with Rapp \cite{rap} is satisfying since he employed a
many-body approach, while we attacked the same physics by using data 
to construct the scattering amplitudes. Nevertheless it should be borne 
in mind that there are effects that are not naturally included here. For 
example, it has been suggested, using effective Lagrangians, that a large, 
negative mass shift can arise from vacuum polarization \cite{jean,zs} or 
quark structure \cite{saito} effects.

This work was supported in part by the US Department of Energy grant 
DE-FG02-87ER40328.


\begin{thebibliography} {30}
\bibitem{rev} C. Adami and G.E. Brown, Phys. Rep. {\bf234}, 1 (1993); 
T.D. Cohen, R.J. Furnstahl, D.K. Griegel and X. Jin, 
Prog. Part. Nucl. Phys. {\bf35}, 221 (1995); 
T.Hatsuda, H. Shiomi and H. Kuwabara, Prog. Theor. 
Phys. {\bf95}, 1009 (1996);
R. Rapp and J. Wambach, in {\it Advances in 
Nuclear Physics}, edited by J.W. Negele and E. Vogt (Plenum, New York, 2000)
Vol. 25, p. 1.
\bibitem{us} V.L. Eletsky, M. Belkacem, P.J. Ellis and J.I. Kapusta,
Phys. Rev. C {\bf64}, 035202 (2001).
\bibitem{je} S. Jeon and P.J. Ellis, 
Phys. Rev. D {\bf 58}, 045013 (1998).
\bibitem{rap} R. Rapp, Phys. Rev. C {\bf63}, 054907 (2001).
\bibitem{man} D.M. Manley and E.M. Saleski, 
Phys. Rev. D {\bf45}, 4002 (1992).
\bibitem{pm} V. Shklyar, G. Penner and U. Mosel, hep-ph/0301152.
\bibitem{har} H. Harari, Phys. Rev. Lett. {\bf20}, 1395 (1968).
\bibitem{col} P.D.B. Collins, {\it An Introduction to Regge Theory and High 
Energy Physics} (Cambridge University Press, Cambridge, UK, 1977).
\bibitem{lee} T.P. Vrana, S.A. Dytman and T.-S.H. Lee, 
Phys. Rep. {\bf328}, 181 (2000).
\bibitem{pdg} Particle Data Group: K. Hagiwara {\it et al.}, 
Phys. Rev. D {\bf 66}, 010001 (2002).
\bibitem{don} A. Donnachie and P.V. Landshoff, 
Phys. Lett. {\bf B296}, 227 (1992).
\bibitem{sib} A. Sibirtsev, C. Elster and J. Speth, nucl-th/0203044.
\bibitem{fol} K. J. Foley {\it et al.}, Phys. Rev.  {\bf181}, 1775 (1969).
\bibitem{ele} V.L. Eletsky, B.L. Ioffe and J.I. Kapusta, 
Eur. J. Phys. A {\bf 3}, 381 (1998); 
V.L. Eletsky and J.I. Kapusta,
Phys. Rev. C {\bf59}, 2757 (1999).
\bibitem{post} M. Post and U. Mosel, Nucl. Phys. {\bf A688}, 808 (2001).
\bibitem{fri} M.F.M. Lutz, G. Wolf and B. Friman, 
Nucl. Phys. {\bf A706}, 431 (2002).
\bibitem{muh} P. M\"uhlich, T. Falter and U. Mosel, nucl-th/0402039.
\bibitem{kling} F. Klingl, T. Waas and W. Weise,
Nucl. Phys. {\bf A650}, 299 (1999).
\bibitem{zsch} S. Zschocke, O.P. Pavlenko and B. K\"ampfer,
Eur. Phys. J. A {\bf 15}, 529 (2002).
\bibitem{ei} V.L. Eletsky and B.L. Ioffe, 
Phys. Rev. Lett. {\bf 78}, 1010 (1997).
\bibitem{mal} S. Mallik and A. Nyffeler, Phys. Rev. C {\bf 63}, 065204 (2001).
\bibitem{sch} R.A. Schneider and W. Weise, Phys. Lett. B {\bf 515}, 89 (2001).
\bibitem{hag} K. Haglin, Nucl. Phys. {\bf A584}, 719 (1995).
\bibitem{alam}
J.-e. Alam, S. Sarkar, P. Roy, B. Dutta-Roy and B. Sinha,
Phys. Rev. C {\bf 59}, 905 (1999).
\bibitem{riek} F. Riek and J. Knoll, nucl-th/0402090.
\bibitem{mt} L.D. McLerran and T. Toimela, Phys. Rev. D {\bf 31},
545 (1985).
\bibitem{w} H.A. Weldon, Phys. Rev. D {\bf 42}, 2384 (1990).
\bibitem{gs} G.J. Gounaris and J.J. Sakurai, Phys. Rev. Lett.
{\bf 21}, 244 (1968).
\bibitem{gk} C. Gale and J.I. Kapusta, Nucl. Phys. {\bf B357}, 65 (1991).
\bibitem{jean} H.-C. Jean, J. Piekarewicz and A.G. Williams,
Phys. Rev. C {\bf 49}, 1981 (1994).
\bibitem{zs} A. Mishra, J. Reinhardt, H. St\"ocker and W. Greiner,
Phys. Rev. C {\bf 66}, 064902 (2002);
D. Zschiesche, A. Mishra, S. Schramm, H. St\"ocker and W. Greiner,
nucl-th/0302073.
\bibitem{saito} K. Saito, K. Tsushima and A.W. Thomas,
Phys. Rev. C {\bf 55}, 2637 (1997).
\end{thebibliography}
\end{document}